# Conduction mechanisms of epitaxial EuTiO$_3$ thin films

R. Zhao,[1] W. W. Li,[1] L. Chen,[2] Q. Q. Meng,[1] J. Yang,[1] H. Wang,[2] Y. Q. Wang,[3] R. J. Tang,[1, a] and H. Yang[1, b]

[1]*Jiangsu Key Laboratory of Thin Films, School of Physical Science and Technology, Soochow University, Suzhou 215006, China*

[2]*Department of Electrical and Computer Engineering, Texas A&M University, College Station, Texas 77843-3128, USA*

[3]*Materials Science and Technology Division, Los Alamos National Laboratory, Los Alamos, New Mexico 87545, USA*

To investigate leakage current density versus electric field characteristics, epitaxial EuTiO$_3$ thin films were deposited on (001) SrTiO$_3$ substrates by pulsed laser deposition and were post-annealed in a reducing atmosphere. This investigation found that conduction mechanisms are strongly related to temperature and voltage polarity. It was determined that from 50 to 150 K the dominant conduction mechanism was a space-charge-limited current under both negative and positive biases. From 200 to 300 K, the conduction mechanism shows Schottky emission and Fowler-Nordheim tunneling behaviors for the negative and positive biases, respectively. This work demonstrates that Eu$^{3+}$ is one source of leakage current in EuTiO$_3$ thin films.

a) Electronic mail: tangrj@suda.edu.cn
b) Electronic mail: yanghao@suda.edu.cn

Multiferroics have cultivated great interest in recent years because of their importance in fundamental research and their diverse applications in spintronics, information storage, and communications.[1,2] However, the scarcity of multiferroic materials remains a great challenge. Many theoretical and experimental approaches have been developed to create new multiferroics.[3] Recently, researchers have found that strain is a simple and effective parameter with which to create multiferroics based on the coupling of spins, optical phonons, and strain.[4,5] The targeted material, EuTiO$_3$ (ETO), was predicted to be a magnetization-electrical polarization (*M-P*) coupled multiferroic with strong and simultaneous ferromagnetism and ferroelectricity (spontaneous magnetization is ~7 Bohr magnetons per Eu, whereas spontaneous polarization is ~10 μC/cm$^2$).[4] These values are comparable to materials that are solely ferroelectric or ferromagnetic. Furthermore, Lee et al. fabricated ETO thin films on DyScO$_3$ substrates (tensile strain ≈+1.1%), and have observed the ferromagnetic and ferroelectric properties of these films with Curie temperatures of 4.24 K and 250 K, respectively.[6] These results not only confirmed the theoretical mechanism but also presented a possible methodology with which to realize a strong ferromagnetic ferroelectric. Leakage issues, however, limited the observation of the ferroelectric hysteresis loops , with no obvious remnant polarization reported directly.

To enhance ferroelectricity and realize strong *M-P* coupling, it is important to investigate the conduction mechanisms and determine how to reduce the leakage current density in the ETO thin films. Although the structural, magnetic, and electric properties of ETO films have been investigated, few reports have addressed conduction mechanisms.[7-10] To determine accurate conduction mechanisms, it is possible to ascertain the temperature dependence of leakage current density-electric field characteristics because it has been proven that the conduction mechanisms of ferroelectric thin films are a strong function of temperature. In this work, we report the temperature-dependent conduction mechanisms of epitaxial ETO thin films and demonstrate that different conduction behaviors are dominant at different temperatures and bias voltages.

Pulsed laser deposition was used to fabricate ETO thin films on (001) $SrTiO_3$ (STO) and (001) Nb-doped $SrTiO_3$ (Nb-STO) substrates. A sintered $EuTiO_3$ ceramic target was focused by a pulsed excimer laser (Lambda Physik, 248 nm, 3 Hz, 2 $J/cm^2$). Deposition temperature was 650℃ and oxygen pressure was $1\times10^{-4}$ Pa. After deposition, the thin films were post-annealed at 1000℃ under a flowing gas of 95vol%Ar + 5vol%$H_2$ for 8.5 hrs to relax the out-of-plane lattice strain and reduce the amounts of $Eu^{3+}$ (discussed later).[7] The crystal structure was confirmed by Rigaku K/Max X-ray diffraction (XRD) and transmission electron microscopy (TEM). The thickness was revealed by cross-sectional TEM. The surface morphology of the thin films was measured by atomic force microscope (AFM) at Asylum Research MFP-3D-SA. Magnetic measurements were performed by a physical properties measurement system (PPMS-9, Quantum Design), with the valence states of the films confirmed by X-ray photoemission spectroscopy (XPS) at PHI5000 VersaProbe. To measure film composition, we used Rutherford backscattering spectrometry (RBS) that consisted of a 3.043-MeV $^4He^+$ beam on a 3-MV Pelletron Tandem accelerator.

Film Eu/Ti ratios were determined by fitting experimental data with commercial Rutherford Universal Manipulation Program (RUMP) software.[11] For electrical measurements, thin films with a thickness of 150 nm were used. Top Pt electrodes with area of $6\times10^{-4}$ cm$^2$ were fabricated by sputtering. Current-voltage characteristics were measured by a Cryogenic probe station (Lakeshore, CRX-4K) with a Keithley 4200-SCS semiconductor parameter analyzer. Bias polarity was defined as positive or negative based on the positive or negative voltage applied to the Pt electrode.

The microstructure of the as-deposited and annealed thin films has been revealed by XRD $\theta$-$2\theta$ scan (not shown). The ETO's lattice parameter is similar to the STO.[6] However, it was determined that the ETO's out-of-plane lattice constant was increased in the as-deposited film. This can be correlated with growth-related strain, as high-vacuum was used during the thin film deposition.[12] Post-annealing has been approved to be effective to relax this strain.[7,13] We found that peaks from the annealed ETO thin film are superimposed on those from the STO substrate. TEM analysis was performed to verify the microstructure of the annealed film. Figure 1 shows (a) the cross-sectional TEM and (b) the high-resolution TEM (HRTEM) images of the annealed ETO on the STO substrate. Figure 1(a) shows that the film has grown as a smooth layer with no obvious grain boundaries in the view area. The inset in Fig. 1(a) shows the corresponded selected-area diffraction pattern, which confirms the high epitaxial quality of the film evidenced by the distinguished diffraction dots from the film/substrate. The HRTEM along the film/substrate interface shows a clean and sharp interface between the film and the substrate without any obvious interfacial reaction. Because the film and the substrate are a perfect match, there is no obvious strain contour along the interface; lattices are continuous and straight from the substrate region to the film region.

AFM measurements of the annealed ETO films' surface morphology (not shown) indicated a roughness of 0.43nm (in term of $R_{rms}$). RBS was used to determine the elemental composition of the films (not shown), revealing a Eu:Ti ratio of ~1:1 for both the as-deposited and annealed ETO films. XPS measurements (Fig. 2) show a Eu 4$d$ core-level photoemission spectrum of as-deposited and annealed ETO thin films with peak fittings. According to previous reports on Eu 4$d$ spectra of europium compounds, fitting peaks at binding energies of 127 eV and 132 eV correspond to $Eu^{2+}$, with peaks at 135 eV and 142 eV corresponding to $Eu^{3+}$.[14,15] We observed that the as-deposited thin films display the existence of trivalent Eu, although the fraction of the remaining $Eu^{3+}$ in the annealed films can be ignored (less than 2%). These results indicate that post-annealing has effectively reduced the $Eu^{3+}$ impurities. It should be noticed that XPS has also been used to detect the valence state of Ti (not shown). No obvious $Ti^{3+}$ impurities have been found for both the as-deposited and annealed ETO films. The annealed films show a distinct antiferromagnetic transition at ~5.0 K measured by PPMS (not shown), which is consistent with the reported value of bulk ETO.[16] All these results show that we have a strain-free, epitaxial, and stoichiometric ETO thin film.

It is possible that impurities are the main source of high leakage current in ferroelectric thin films.[17] It is expected that the $Eu^{3+}$ in ETO thin films contributes to the leakage current. A comparison of leakage current density of the as-deposited and the annealed ETO films at room temperature shows that the leakage current density at 50 kV/cm has been reduced by seven orders of magnitude as a result of post-annealing (see inset of Fig. 3). Combined with the above XPS results, we conclude that the $Eu^{3+}$ is one possible origin of leakage current in ETO films. The possible mechanism for such leakage is that the impurity of $Eu^{3+}$—its increase cation valence—contributes to more charge carriers that in turn lead to higher leakage current density. This mechanism makes it possible to decrease effectively the leakage current density by reducing the amount of $Eu^{3+}$.

Obtaining further details of the leakage current requires the analysis of the leakage current density *vs.* electric field (*J-E*) characteristics of the annealed ETO thin films. Figure 3 shows the typical temperature dependence of the *J-E* characteristics in annealed ETO thin films for temperatures that range from 50 to 300 K. The leakage current density is approximately $10^{-4}$ A/cm$^2$ at 300 kV/cm and 300 K, units comparable to or lower than those of the ferroelectric $BiFeO_3$ and $(Ba,Sr)TiO_3$ thin films.[18,19] Four conduction mechanisms have been included to investigate leakage behaviors in ferroelectric thin films: Schottky emission, Poole-Frenkel (PF) emission, Fowler-Nordheim (FN) tunneling, and space-charge-limited current (SCLC). PF and SCLC consist of bulk-limited conduction, compared with the interface-limited conduction of the Schottky emission and the FN tunneling. From 200 to 300 K, *J-E* curves asymmetrically at positive and negative biases, a phenomenon caused by the different work function of the Pt and Nb-STO electrodes that could be correlated with the interface-limited conduction process. However, the *J-E* curves are symmetric from 50 to 150 K, a phenomenon believed to result from the bulk-limited conduction process. Leakage currents are mainly controlled by Ohmic contact in the low-electric-field range at both negative and positive biases for the whole temperature range. We point out that we considered all four mechanisms in the present work, but only the proper mechanisms are discussed in more detail below.

Originally derived from the metal/semiconductor interface, the Schottky equation is widely applied to the leakage behavior at high electric fields:[20]

$$J = A^* T^2 \exp\left[\frac{-q(\phi_b - \sqrt{qE/4\pi\varepsilon_i\varepsilon_0})}{kT}\right], \qquad (1)$$

where *J* is the current density, $A^*$ is the Richardson constant, *T* is the temperature, $\varepsilon_i$ is the optical dielectric constant, $\varepsilon_0$ is the permittivity of free space, *E* is the applied electric field, *k* is the Boltzmann's constant, and $q\phi_b$ is the Schottky barrier height. As is typical when applying the Schottky emission, a linear relation of $\ln(J/T^2)$ and $E^{1/2}$ should be observed, with the slope giving the $\varepsilon_i$. Figure 4(a) shows the $\ln(J/T^2)$ *vs.* $E^{1/2}$ characteristics of the annealed ETO films at negative bias. At high electric fields (>300kV/cm), the plots consist of straight lines for the whole temperature range. The calculated $\varepsilon_i$ (shown in the figure) decreases with increasing temperature. The optical refractive index (*n*) can be determined from the $\varepsilon_i$ with a relationship of $n = \sqrt{\varepsilon_i}$ (shown in the figure). On the other hand, spectroscopic ellipsometry (SE) measurements by Lee et al determined an *n* value of 1.88 at the wavelength of 633 nm.[8] The *n* values calculated from the slopes are in good agreement with the SE-extracted results in temperatures that range from 200 to 300 K. Therefore, leakage behavior at negative bias is controlled by the Schottky emission at high electric fields (>300 kV/cm) from 200 to 300 K.

FN tunneling can be used to tunnel through an interfacial energy barrier to inject charge carriers from electrodes into the insulator layer:[20]

$$J = BE^2 \exp\left(\frac{-C\phi_i^{3/2}}{E}\right), \qquad (2)$$

where $B$ and $C$ are constants and $\phi_i$ is the potential barrier. In the case of FN tunneling, a linear relationship between $\ln(J/E^2)$ and $(1/E)$ should be observed. Figure 4(b) shows the $\ln(J/E^2)$ and $(1/E)$ characteristics of the annealed ETO films at positive bias for temperatures that range from 200 to 300 K. Because FN tunneling is a thermally assisted tunneling process, it is easy to determine FN tunneling behavior at high electric fields by observing the leakage currents, where the onset electric field ($E_t$) decreases with increasing temperature.[21]

Hall measurements taken by Takahashi et al. showed that electrons serve as main charge carriers in ETO films.[22] At positive bias, electrons are injected from the bottom electrode (Nb-STO) to the ETO layer. The depletion layer at the interface should be thin because ETO and Nb-STO are both oxides. It is reasonable to assume that the leakage current shows FN tunneling behavior at the positive bias. Considered as a common conduction mechanism in ferroelectric oxide materials, FN tunneling has also been found on thin films such as $BiFeO_3$ and $(Ba_{0.5}Sr_{0.5})TiO_3$.[23] Based on this knowledge, it is clear that from 200 to 300 K the dominant conduction mechanisms are the Schottky emission and FN tunneling for negative and positive biases, respectively. Both are interface-limited process, which is consistent with the above discussion.

Figure 4(c) shows the log($J$) *vs.* log($E$) characteristics of the annealed ETO films at negative bias from 50 to 150 K. The plots show a linear relation with different slopes at two electric-field regions and indicate a power law relationship of $J \propto E^\alpha$, which is a key characteristic of SCLC.[22] When charge carrier density becomes greater than that of thermally simulated charge carriers (as a result of carrier injection), the SCLC should dominate the leakage behavior, as expressed in the following:[24]

$$J = \frac{9}{8}\varepsilon_r\varepsilon_0\mu\theta\frac{E^2}{d}, \qquad (3)$$

where $\theta$ is the ratio of free-to-trapped charge carriers, $\mu$ is the mobility of charge carriers, $\varepsilon_r$ is the relative dielectric constant, and $d$ is the film thickness. As shown in Fig. 4(c), the slopes are close to 2 (the middle electric-field range), results that agree well with Eq. (3). Increasing the electric field changes the slopes to ~4.5, possibly because the electric field is close to a trap-filled limit, causing the slopes to deviate from 2.[20,21] A log($J$) *vs.* log($E$) analysis at positive bias at these temperatures also yields similar results (not shown). Therefore, we have determined that SCLC dominates leakage behavior at both positive and negative biases from 50 to 150 K.

To summarize, we investigated the temperature dependence of the *J-E* characteristics in ETO thin films. From 50 to 150 K, the leakage current was found to be controlled by bulk-limited SCLC behavior for both positive and negative biases. From 200 to 300 K, the dominant mechanism changes to an interface-limited mechanism as a result of Schottky emission (negative bias) and FN tunneling (positive bias). We also demonstrated that post-annealing in the reducing atmosphere effectively decreases the leakage current by reducing $Eu^{3+}$ impurities.

The authors acknowledge the support of the National Natural Science Foundation of China under Grant No. 11004145, the Natural Science Foundation of Jiangsu Province under Grant No. SBK201021263, and the Scientific Research Foundation for the Returned Overseas Chinese Scholars (State Education Ministry of China). Partial support for thin film characterization was also provided by the Center for Integrated Nanotechnologies, a DOE nanoscience user facility, jointly operated by Los Alamos and Sandia National Laboratories. The TEM work at Texas A&M University is funded by the US National Science Foundation (NSF-0846504).

**Figures**

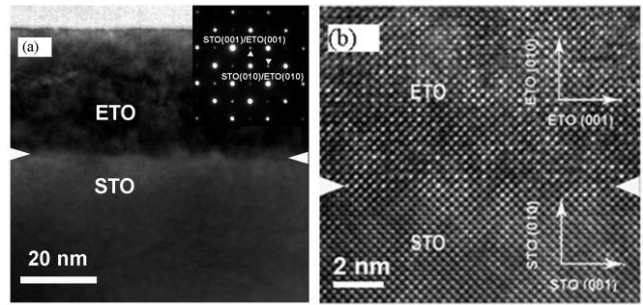

Figure 1. ETO thin films as imaged by (a) cross-section TEM and (b) HRTEM. The inset in Fig. 1 (a) shows a corresponded selected-area diffraction image.

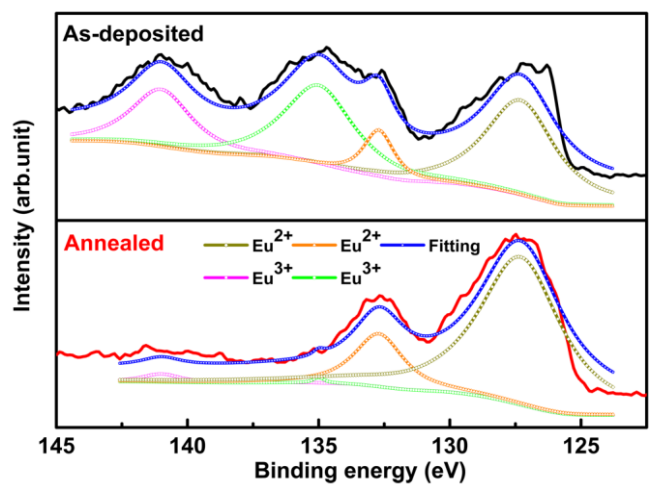

Figure 2. The open squares represent peak fittings on these graphs, which show the Eu 4*d* core-level photoemission spectra of as-deposited and annealed ETO thin films.

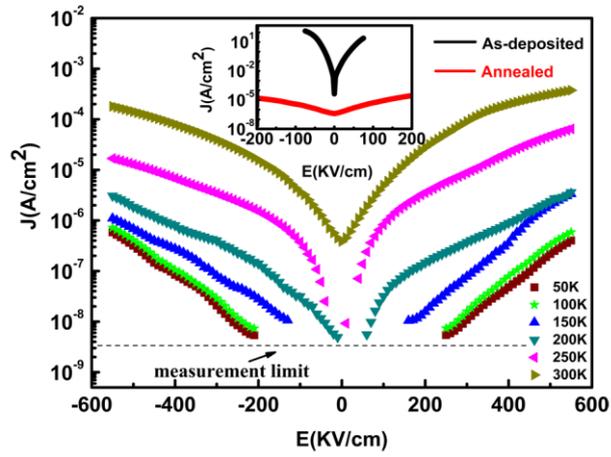

Figure 3. This graph shows the typical temperature dependence of *J-E* characteristics of annealed ETO thin films in temperatures that range from 50 to 300 K. The dotted line shows the measurement limit. The inset shows a comparison of leakage current density between the as-deposited and the annealed ETO films at room temperature.

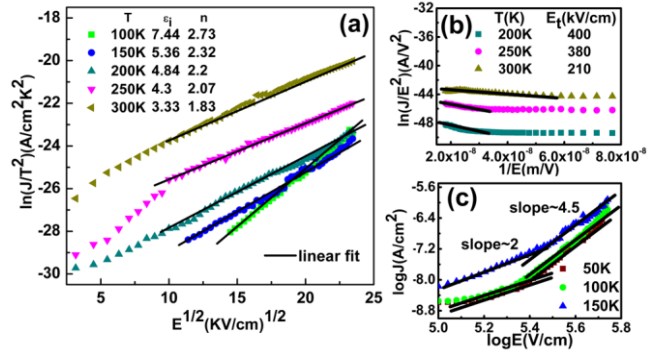

Figure 4. (a) This graph shows ln($J/T^2$) *vs.* $E^{1/2}$ characteristics of the annealed ETO films at negative bias. Listed on the graph are the calculated optical dielectric constant ($\varepsilon_i$) and optical refractive index (*n*). (b) This graph shows the ln($J/E^2$) and (1/$E$) characteristics of the annealed ETO films at positive bias from 200 to 300 K. The onset electric field of the FN tunneling ($E_t$) decreases with increasing temperature. (c) This graph shows log(*J*) *vs.* log(*E*) characteristics of the annealed ETO films at negative bias from 50 to 150 K.